# Value of Information Analysis via Active Learning and Knowledge Sharing in Error-Controlled Adaptive Kriging

Chi Zhang[1], Zeyu Wang[1], and Abdollah Shafieezadeh[1]
[1]Risk Assessment and Management of Structural and Infrastructure Systems (RAMSIS) Lab, Department of Civil Environmental and Geodetic Engineering, The Ohio State University, Columbus, OH 43210 USA

Corresponding author: Abdollah Shafieezadeh (e-mail: Shafieezadeh.1@ osu.edu).

Large uncertainties in many phenomena have challenged decision making. Collecting additional information to better characterize reducible uncertainties is among decision alternatives. Value of information (VoI) analysis is a mathematical decision framework that quantifies expected potential benefits of new data and assists with optimal allocation of resources for information collection. However, analysis of VoI is computational very costly because of the underlying Bayesian inference especially for equality-type information. This paper proposes the first surrogate-based framework for VoI analysis. Instead of modeling the limit state functions describing events of interest for decision making, which is commonly pursued in surrogate model-based reliability methods, the proposed framework models system responses. This approach affords sharing equality-type information from observations among surrogate models to update likelihoods of multiple events of interest. Moreover, two knowledge sharing schemes called model and training points sharing are proposed to most effectively take advantage of the knowledge offered by costly model evaluations. Both schemes are integrated with an error rate-based adaptive training approach to efficiently generate accurate Kriging surrogate models. The proposed VoI analysis framework is applied for an optimal decision-making problem involving load testing of a truss bridge. While state-of-the-art methods based on importance sampling and adaptive Kriging Monte Carlo simulation are unable to solve this problem, the proposed method is shown to offer accurate and robust estimates of VoI with a limited number of model evaluations. Therefore, the proposed method facilitates the application of VoI for complex decision problems.

## I. INTRODUCTION
Real-world phenomena are often accompanied with large uncertainties that may have aleatory or epistemic nature. Prominent examples are risks posed by natural or manmade hazards to the built, natural, or human systems. Such hazards threaten the functionality and even the integrity of systems; thus, requiring actions that reduce the risk to the society. To reduce the impact of such disturbances, a common strategy is to allocate budget for actions that enhance the performance of systems under uncertainty. Ideally, with more budget spent, the reliability of systems increases. However, limitations in budget warrant careful considerations, highlighting the importance of allocating limited budgets in a cost-effective fashion. Risk-informed decision-making methods offer an approach to analyze risks and examine the effectiveness of a set of decision alternatives to identify appropriate strategies. The presence of high uncertainties in various elements that contribute to risk, however, poses a significant challenge for decision making. An often neglected alternative to taking actions that directly enhance the physical or operational performance of systems is to collect more information about the system or disturbances in order to better inform the decisions. However, obtaining new information via, for example, field observations, lab or field experiments, or high-fidelity simulations is costly and sometimes faces feasibility challenges. Consequently, among key questions are whether collection of new information is beneficial to risk management, for what uncertain risk contributors collection of new information is effective, and how much new information with what fidelity is optimal. One can perform an analysis of extant uncertainty in variables and models, and present risk contributing factors with highest uncertainty as those that warrant collecting more information. However, such an analysis misses the fact that while a factor may have high uncertainty, its contribution to risk may not be significant. Therefore, collecting more information may have negligible effects on risk estimates. For critical systems such as nuclear power plants, importance measures based on probabilistic safety assessments have been developed [1]. Risk importance measures, as the name implies, can quantify the degree of contribution of various involved components to risk, which can then be used to identify the most important risk factors. Different risk importance measures have been proposed to serve different purposes. For example, Fussell-Vesely (FV) importance [2], [3] is often used as a risk reduction indicator, and represents the maximum decrease in risk for an improvement in an element that corresponds to the basic event. Another metric is Risk Achievement Worth (RAW) [4], which is a measure of the worth of the basic event in achieving the present level of risk and the importance of maintaining the current level of reliability for the basic event. Birnbaum Importance (BI) [5] is commonly used when the risk measure is defined to be the system unavailability or unreliability [1]. More recently, Mandelli et al. [6] proposed a mathematical framework for determining risk importance measures in a dynamic probabilistic risk assessment framework, in which classical measures (e.g., FV and RAW) can be determined from simulation-based data instead of minimal cut sets. These methods can be used to prioritize the information collection process. However, the ultimate question vis-à-vis collection of additional data is whether new information can impact decisions. As an example, let's consider the case of a nuclear power plant (NPP) that is planned for license renewal. Following the Tohoku earthquake and tsunami in 2011 that led to the meltdown of several reactors in Fukushima Daiichi nuclear power plant [7], some provisions for safety assessment of NPPs in the US were changed. These new requirements must be considered in applications of NPPs for license renewal. In the face of more strict requirements, NPP owners would desire the least costly strategy that satisfies the safety requirements. In such a circumstance, new information, even if it has notable impacts on risk estimates, is valuable only if it leads to a change in the decision, and the savings offered by the new information exceed the associated costs of data collection. The uncertainty analysis methods reviewed above are not able to address this significant need.

Value of Information (VoI) analysis, which was originally introduced in [8]–[10], facilitates this process through estimation of the benefit of new information. VoI analysis uses Bayesian inference to update prior probabilities with potential evidences to produce an estimate of posterior probabilities. VoI determines expected losses due to errors in decision making because of present uncertainties, and identifies the best strategy for collection of new information such that the greatest net benefit to the decision maker is achieved. The application of VoI can be found in several science and engineering fields. In applied ecology, VoI analysis has been used to quantify how additional information may improve management outcomes [11]. In health economics, Heath et al. [12] used VoI analysis to quantify the expected value of obtaining perfect information about the underlying model parameters, placing an upper bound on the cost of additional research aimed at reducing uncertainty. In prognostics and health management, Heckerman et al. [13] computed the value of information of sets of tests to provide a nonmyopic, yet tractable alternative to traditional myopic analyses for identifying the next best piece of evidence to observe. In military communications, Suri et al. [14] explored VoI-based approaches to prioritize and filter information disseminated to and from dismounted soldiers over tactical networks. In reinforcement learning, Sledge et al. [15] used VoI to provide the optimal tradeoff between the expected return of a policy and the policy's complexity.

Despite all the benefits, VoI analysis can be computationally very challenging. By its nature, VoI analysis requires an estimation of posterior probabilities for the information to be collected in the future. When the information is of the equality type, which is the case for most information, a direct evaluation of the posterior probability is usually not available [16]. Hence, simulation-based methods, such as Monte Carlo Simulation (MCS) are often used in the process of the analysis. These methods usually require a large number of model evaluations. For cases involving complex, high-fidelity simulations, each such simulation is costly and can take a long time even on supercomputers. Straub [17] proposed a framework for the computation of VoI based on structural reliability methods to analyze event probabilities. In that framework, Importance Sampling was adopted to greatly improve the efficiency while maintaining the accuracy of VoI analysis. In structural reliability, the construction of importance sampling density is often based on the most likely failure points (MLFPs). For the identification of these points, techniques such as First Order Reliability Method (FORM) or Secord Order Reliability Method (SORM) [18], [19] are often used. Although FORM and SORM are widely popular, they face several limitations. The MLFPs are imaginary points and sometimes are difficult to locate. Moreover, the accuracy of these methods may not be satisfactory for high-dimensional and highly nonlinear problems.

To address the computational challenge of VoI analysis, this paper proposes a novel approach based on adaptive Kriging surrogate modeling. Polynomial Response Surface [20]–[22], Polynomial Chaos Expansion (PCE) [23], Support Vector Regression (SVR) [24], [25], and Kriging [26] are among popular surrogate models. However in recent years, Kriging has gained significant attention for its ability to provide estimates of expected model response and the associated variance [27]–[33]. Kriging-based reliability methods such as Efficient Global Reliability Analysis (EGRA) [34] and Active learning reliability method combining Kriging and Monte Carlo Simulation (AK-MCS) [26] have leveraged these properties to identify the next best training points and adaptively construct surrogate models. These techniques, however, are shown to require unnecessary calls to costly performance functions or converge to inaccurate estimates of event probabilities [35]. The proposed framework for VoI analysis integrates a variant of Kriging-based reliability analysis method proposed by the authors [35], which is capable of establishing an upper bound for error in reliability estimates. This feature is leveraged here to strategically train surrogate models to achieve desired accuracy with minimum computational demand.

In many applications, risks to a system may be comprised of several potential events each with a likelihood of occurrence and a set of consequences. Limit states describing the events of interest associated with a phenomenon (e.g., different levels of damage to a system under an extreme event) are often described by establishing different thresholds for a performance metric. There can be cases where multiple performance metrics need to be considered, where each performance metric has multiple thresholds. One approach to analyze such decision problems is to establish a surrogate model for each limit state i.e., each pair of performance metric and threshold. This method, however, requires a large number of calls to performance functions which can be computationally demanding especially if it involves expensive computational models. For such problems instead of modeling the limit state functions, which is commonly pursued in surrogate model-based reliability methods, the proposed framework in this research models system responses. This approach affords sharing equality-type information from observations among surrogate models to update likelihoods of multiple events of interest. Two knowledge sharing schemes are proposed to take most advantage of costly model evaluations. One scheme lets the analysis process share knowledge between groups by passing the training points, and the other directly construct shared models for multiple limit states to share knowledge to the greatest degree. These knowledge sharing methods further reduce the computational cost of VoI analysis for cases involving multiple potential actions for several limit state functions each describing the onset of a key event that may incur cost. The accuracy and efficiency of the proposed method are investigated for an example involving monitoring of a truss bridge.

## II. VALUE OF INFORMATION ANALYSIS

Information collection of a decision problem is worthy if the associated VoI minus the cost of collecting the information is larger than zero. The plan for information collection is optimal when the above quantity is maximum. A detailed computational procedure of VoI analysis based on structural reliability methods is presented by Straub in [17]. The proposed method here builds on this approach to enable the analysis of VoI for complex problems and substantially improve the computational efficiency. The VoI analysis method in [17] will be briefly introduced in the rest of this section.

As noted, VoI compares the minimum cost that can be achieved with prior information to that using the yet unknown future information. Therefore, one of the elements of VoI analysis requires optimization of decisions based on prior information. For decision making purposes, discrete events are often used to denote the system state, such as failure F and survival S. Note that in some cases failure can contain different damage states. Let $X$ denote the vector containing involved random variables. The discrete events that represent the different system states correspond to domains in the outcome space of $X$. Furthermore, let $E_1, E_2, \ldots, E_m$ denote the mutually exclusive and collectively exhaustive set of the discrete events. For example, an event $E_s$ can be defined using a limit state function $g_s(X)$ corresponding to state $s$ as $E_s = \{g_s(X) \leq 0\}$. The decision optimization based on prior information can be formulated as follows:

$$a_{opt} = arg \min_a \sum_{i=1}^{m} c_{E_i}(a) \, Pr(E_i) \tag{1}$$

where $c_{E_i}(a)$ is the cost corresponding to event $E_i$ and decision $a$, $Pr(E_i)$ is the probability of event $E_i$, which can be calculated using structural reliability methods. Subsequently, the minimum expected cost can be formulated as follows:

$$C_{prior} = \min_a \sum_{i=1}^{m} c_{E_i}(a) \, Pr(E_i) \tag{2}$$

Additional information can potentially afford reaching better decisions. Let's first assume that the new information is perfect, i.e. following the collection of new information, there is no uncertainty in $X$. Thus, one can take the action that minimizes the cost based on this 'perfect information'. The optimal action for a given $x$ can be formulated as:

$$a_{opt}^*(x) = arc \min_a c(a, x) \tag{3}$$

The conditional value of perfect information ($CVoPI$) given $X = x$ can be written as the difference between the cost associated with $a_{opt}$ and the cost of $a_{opt}^*$:

$$CVoPI(x) = c(a_{opt}, x) - c(a_{opt}^*, x) \tag{4}$$

where $c(a, x)$ is the cost corresponding to decision $a$ given $X = x$. $CVoPI$ can also be formulated for a given event $E_i$ as follows:

$$CVoPI_{E_i} = c_{E_i}(a_{opt}) - c_{E_i}(a_{opt,i}^*) \tag{5}$$

where $a_{opt,i}^*$ is the optimal solution given that event $E_i$ occurs. The value of perfect information ($VoPI$) can subsequently be calculated as the expected value of the $CVoPI$ as follows:

$$VoPI = C_{prior} - \sum_{i=1}^{m} c_{E_i}(a_{opt,i}^*) Pr(E_i) \tag{6}$$

Here, $VoPI$ represents the upper bound of the value that any information collection approach can offer. In real world, perfect information is not available as measurements are commonly subject to extrinsic and intrinsic errors. In addition, it is not practical or sometimes even feasible to collect information on all $X$. Therefore, only imperfect information is available for most cases. Albeit imperfect, it can still provide valuable knowledge about the probabilistic properties of $X$ and the probabilities of $E_1, E_2, \ldots, E_m$. Bayesian updating is a mathematical framework used for this purpose.

For a given piece of new information $Z$, which can include direct measurements of random variables in $X$ or the system response, the updated probability of event $E_i$, $Pr(E_i|Z)$, can be estimated using Monte Carlo simulation as follows:

$$\Pr(E_i|Z) \approx \frac{\sum_{k=1}^{n_{MCS}} I(x_k \in \Omega_{E_i}) L(x_k)}{\sum_{k=1}^{n_{MCS}} L(x_k)} \tag{7}$$

where $x_k, k = 1, \ldots, n_{MCS}$, are samples from the joint probability density functions $f_X(x)$ using MCS, $I(x_k \in \Omega_{E_i})$ is the indicator function that indicates whether $x_k \in \Omega_{E_i}$ and the likelihood function $L(x_k)$ represents the relation between the random variable $X$ and the information $Z$. Note that this paper is focused on the information of equality type, as it is the most common type and the most difficult to deal with. Let $q(X) = [q_1(X), \ldots, q_K(X)]$ denote the observation quantities and $y = [y_1, \ldots, y_k]$ denote the corresponding measurements of the quantities. Assume $y_i$ has an additive error from modeled by the statistically independent random variable $\varepsilon_i$. Thus, the following relationship holds: $y_i - q_i(x) = \varepsilon_i$. Accordingly, the likelihood function of the data collection outcome can be formulated as follows:

$$L(x) = f_Y(y|x) = \prod_{i=1}^{m} f_{\varepsilon_i}[y_i - q_i(x)] \tag{8}$$

Similar to (1), the optimal action for the decision optimization problem given an observation $Z$, i.e., posterior decision analysis, can be solved using the following equation:

$$a_{opt|Z} = arg\ \min_{a} \sum_{i=1}^{m} c_{E_i}(a)\ Pr(E_i|Z) \tag{9}$$

Correspondingly, the conditional value of information given $Z$ can be computed as follows:

$$CVoI_Z = \sum_{i=1}^{m} c_{E_i}(a_{opt})\ Pr(E_i|Z) - \sum_{i=1}^{m} c_{E_i}(a_{opt|Z})\ Pr(E_i|Z) \tag{10}$$

The expected value of $CVoI_Z$ over all the measurement outcomes is the value of information ($VoI$). For $l$ mutually exclusive domains of measurement outcomes $Z_1, Z_2, ..., Z_l$, the corresponding $VoI$ can be calculated similar to (6) using:

$$VoI = C_{prior} - \left[ \sum_{j=1}^{l} Pr(Z_j) \min_{a} \sum_{i=1}^{m} c_{E_i}(a) Pr(E_i|Z_j) \right] \tag{11}$$

For continuous measurement outcomes $Z = \{Y = y\}$, i.e., the equality information, $VoI$ can be calculated as:

$$VoI = C_{prior} - \left[ \int_{Y} f_Y(y) \min_{a} \sum_{i=1}^{m} c_{E_i}(a) Pr(E_i|Y = y)\,dy \right] \tag{12}$$

where $f_Y(y)$ is the joint PDF of the data collection outcome $Y$. Using Monte Carlo simulation, $VoI$ in (12) can be estimated as follows:

$$VoI \approx C_{prior} - \frac{1}{n_{SY}} \left[ \sum_{j=1}^{n_{SY}} \min_{a} \sum_{i=1}^{m} C_{E_i}(a) \frac{\sum_{k=1}^{n_{MCS}} I(x_k \in \Omega_{E_i}) L(x_k|y_j)}{\sum_{k=1}^{n_{MCS}} L(x_k|y_j)} \right] \tag{13}$$

where $L(x_k|y_j)$ is the likelihood of $x_k$ conditional on the data collection outcome $y_j$, and $x_k, k = 1, ..., n_{MCS}$ and $y_j, j = 1, ..., n_{SY}$ are samples from the joint probability density functions $f_X(x)$ and $f_Y(y)$, respectively. According to [36], $x_k$s are sampled first from $f_X(x)$. Subsequently for each $x_k$, the probability density function $f_{Y|X}(y|x_k)$ can be determined using (8), and $y_j$s are generated from it.

In this approach, $n_{MCS}$ samples are generated for $x_k$ through MCS method. For each $x_k$, one model evaluation is required; therefore, the total number of model evaluations is $n_{MCS}$. As a result, the computational cost of calculating the VoI for problems involving high-fidelity models can be significantly high. To improve the efficiency, Straub [17] proposed an importance sampling (IS)-based scheme to more efficiently sample random variables. Although this approach considerably improves the computational efficiency of VoI analysis, the required number of evaluations of costly models is often still high, rendering the application of VoI not practical. Moreover, IS highly relies on the chosen IS density function, which is constructed using MLFPs. Locating MLFPs requires additional model evaluations, and is often challenging to locate them with sufficient accuracy. Therefore, despite improved efficiency, IS-based schemes may have limited applications. In this paper an alternative approach is proposed based on adaptive surrogate modeling concepts. More specifically, a framework for VoI analysis is proposed in which an adaptive Kriging Monte Carlo simulation with an error rate-based stopping criterion called ESC [35] is integrated to substantially enhance the computational efficiency of VoI analyses. The framework also includes two novel knowledge sharing strategies among surrogate models for the discrete events of interest. As adaptive Kriging is the basis of this framework, Section III describes this surrogate modeling approach. The proposed methodology is then elaborated in detail in Section IV.

### III. ADAPTIVE KRIGING WITH ERROR-BASED STOPPING CRITERION
The main idea behind adaptive Kriging Monte Carlo simulation is to train high-fidelity surrogate models to substitute the original models, which are computationally expensive. A brief overview of Kriging surrogate models is provided in the following subsection.

#### A. KRIGING SURROGATE MODELS
Kriging meta-models have been widely used to replace computationally sophisticated phenomenological models with easy-to-evaluate analytical forms that can properly represent the response of the original models.

Let's consider $g(X_g)$ as the original model with variables $X_g$. The stochastic estimator of $g(X_g)$ based on Kriging shown by $\hat{g}(X_g)$ has the following form:

$$\hat{g}(X_g) = F(\boldsymbol{\beta}, x_g) + \mathcal{GP}(x_g) = \boldsymbol{\beta}^T f(x_g) + \mathcal{GP}(x_g) \tag{14}$$

where $x_g$ is the vector of random variables, $F(\boldsymbol{\beta}, x_g)$ are regression elements, and $\mathcal{GP}(x_g)$ is a Gaussian process. In $F(\boldsymbol{\beta}, x_g)$, $f(x_g)$ is the Kriging basis and $\boldsymbol{\beta}$ is the corresponding set of coefficients. There are multiple formulations of $\boldsymbol{\beta}^T f(x_g)$ including ordinary ($\beta_0$), linear $\left(\beta_0 + \sum_{i=1}^{N} \beta_i x_{g_i}\right)$, or quadratic $\left(\beta_0 + \sum_{i=1}^{N} \beta_i x_{g_i} + \sum_{i=1}^{N} \sum_{j=1}^{i} \beta_{ij} x_{g_i} x_{g_j}\right)$, where $N$ is the number of dimensions of $x_g$. In this article, the ordinary Kriging model is used. The Gaussian process $\mathcal{GP}(x_g)$ has a zero mean and a covariance matrix that can be represented as:

$$COV\left(\mathcal{GP}(x_g^{(i)}), \mathcal{GP}(x_g^{(j)})\right) = \sigma^2 R(x_g^{(i)}, x_g^{(j)}; \boldsymbol{\theta}) \tag{15}$$

where $\sigma^2$ is the process variance or the generalized mean square error (MSE) from the regression, $x_g^{(i)}$ and $x_g^{(j)}$ are two observations, and $R(x_g^{(i)}, x_g^{(j)}; \boldsymbol{\theta})$ is known as the kernel function representing the correlation between observations $x_g^{(i)}$ and $x_g^{(j)}$ parametrized by $\boldsymbol{\theta}$. The correlation functions implemented in Kriging can include, among others, linear, exponential, Gaussian, and Matérn functions. The Gaussian kernel function is used in this paper, which has the following form:

$$R(x_g^{(i)}, x_g^{(j)}; \boldsymbol{\theta}) = \prod_{k=1}^{N} exp\left(-\theta^k (x_{g_k}^{(i)} - x_{g_k}^{(j)})^2\right) \tag{16}$$

where $x_{g_k}^{(i)}$ is the $k_{th}$ dimension of $x_g^{(i)}$ and $\boldsymbol{\theta}$ is estimated via the Maximum Likelihood Estimation (MLE) method [37]. It is shown that the variation of $\boldsymbol{\theta}$ has significant impact on the performance of the Kriging meta-model [38]–[40]. To maintain consistency, $\theta^k$ can be found using the optimization algorithms in DACE [41], [42] or UQLab [37]. The UQLab package and MLE approach are used in this research. The formulation based on MLE is as follows:

$$\boldsymbol{\theta}^* = \underset{\boldsymbol{\theta}}{argmin}\left(\left|R(x_g^{(i)}, x_g^{(j)}; \boldsymbol{\theta})\right|^{\frac{1}{m}} \sigma^2\right) \tag{17}$$

where $m$ is the number of training points. Accordingly, regression coefficients $\boldsymbol{\beta}$, and the predicted mean and variance can be determined as follows [37]:

$$\begin{aligned}\boldsymbol{\beta} &= (F^T R^{-1} F)^{-1} F^T R^{-1} Y \\ \mu_{\hat{g}}(x_g) &= f^T(x_g)\boldsymbol{\beta} + r^T(x_g) R^{-1}(y - F\boldsymbol{\beta}) \\ \sigma_{\hat{g}}^2(x_g) &= \sigma^2 - \sigma^2 r^T(x_g) R^{-1} r(x_g) + \sigma^2 u^T(x_g)(F^T R^{-1} F)^{-1} u(x_g)\end{aligned} \tag{18}$$

where $F$ is the matrix of the basis function $f(x_g)$ evaluated at the training points, i.e., $F_{ij} = B_j(x_g^{(i)})$, $i = 1, 2, \ldots, m; j = 1, 2, \ldots, p$, $r(x_g)$ is the correlation between known training points $x_g^{(i)}$ and other points $x_g$: $r_i = R(x_g, x_g^{(i)}, \boldsymbol{\theta})$, $i = 1, 2 \ldots m$, $R$ is the autocorrelation matrix for known training points: $R_{ij} = R(x_g^{(i)}, x_g^{(j)}, \boldsymbol{\theta})$, $i = 1, 2, \ldots, m; j = 1, 2, \ldots, m$, and $u(x_g) = F^T R^{-1} r(x_g) - f(x_g)$. Therefore, $\hat{g}(x_g)$ can be presented using the estimated Kriging mean $\mu_{\hat{g}}(x_g)$ and variance $\sigma_{\hat{g}}^2(x_g)$ as:

$$\hat{g}(x_g) \sim N\left(\mu_{\hat{g}}(x_g), \sigma_{\hat{g}}^2(x_g)\right) \tag{19}$$

As can be observed, responses from the Kriging model $\hat{g}(x_g)$ are not deterministic but stochastic. The uncertainty information offered by the Kriging model can be used for the identification of next best training points and to develop appropriate criteria to stop training. Wang and Shafieezadeh [35] have proposed an efficient error rate-based stopping criterion for adaptive reliability analysis. This method, which takes advantage of the uncertainty information provided by Kriging, is introduced next.

### *B. AN EFFICIENT ERROR-BASED STOPPING CRITERION*
Note that the stopping criterion for terminating the training procedure is a key step in adaptive Kriging methods. If the stopping criterion is set loosely, the estimated probability of failure may not be accurate. On the other hand, the training process will use a large number of unnecessary trainings and become computationally inefficient if the stopping criterion is overly conservative. To address this challenge, an efficient stopping criterion called ESC was proposed by Wang and Shafieezadeh in [35]. In ESC, the maximum error rate, $\hat{\epsilon}_{max}$, for the estimated probability of failure is derived and treated as a criterion to evaluate the accuracy of the Kriging surrogate model in representing a limit state function. First, the probability of failure estimated through the Kriging surrogate can be written as:

$$\hat{P}_f = \frac{\hat{N}_f}{N_{mcs}} \tag{20}$$

where $\hat{N}_f$ denotes the estimated number of failure points in $S$. The true failure probability based on crude MCS is:

$$P_f = \frac{N_f}{N_{mcs}} \tag{21}$$

where $N_f$ is the true number of failure points. Thus, the relative error of $\hat{P}_f$ can be defined as:

$$\epsilon = \left|\frac{\hat{P}_f}{P_f} - 1\right| = \left|\frac{\hat{N}_f - N_f}{N_f}\right| \tag{22}$$

The estimated failure domain is denoted as $\hat{\Omega}_f$, the safe domain as $\hat{\Omega}_s$, the total number of wrong sign estimations in $\hat{\Omega}_f$ as $\hat{S}_f$, and in $\hat{\Omega}_s$ as $\hat{S}_s$. Note that $\hat{\Omega}_f, \hat{\Omega}_s \in \Omega$, and $\hat{\Omega}_f \cap \hat{\Omega}_s = \emptyset$. In the Kriging model, $N_f, \hat{S}_s$, and $\hat{S}_f$ are not deterministic but follow Poisson binomial distributions as shown in [35]. $N_f$ can therefore be estimated as:

$$N_f = \hat{N}_f + \hat{S}_s - \hat{S}_f \tag{23}$$

Here, both $\hat{S}_s$ and $\hat{S}_f$ follow a Poisson binomial distribution with mean and variance shown below:

$$\hat{S}_s \sim PB\left(\sum_{i=1}^{\hat{N}_s} P_i^{wse}, \sum_{i=1}^{\hat{N}_s} P_i^{wse}(1 - P_i^{wse})\right) \tag{24}$$
$$x_i \in \hat{\Omega}_s$$

$$\hat{S}_f \sim PB\left(\sum_{i=1}^{\hat{N}_f} P_i^{wse}, \sum_{i=1}^{\hat{N}_f} P_i^{wse}(1 - P_i^{wse})\right) \tag{25}$$
$$x_i \in \hat{\Omega}_f$$

where $PB$ denotes the Poison Binomial distribution and $P_i^{wse}$ denotes the probability of wrong sign estimation for $x_i$, which can be computed as $P_i^{wse} = \Phi(-U'(x_i))$, where $\Phi(\cdot)$ is the standard normal cumulative density function and $U'(\cdot)$ is the $U'$ learning function in [26]: $U_i'(x_g) = \frac{|\mu_{\hat{g}}(x_g)|}{\sigma_{\hat{g}}(x_g)}$. Therefore, with a confidence level $\alpha$, the upper and lower bounds of $\hat{S}_s$ and $\hat{S}_f$ can be found as:

$$\hat{S}_s \in \left(\boldsymbol{\Theta}_{\hat{S}_s}^{-1}\left(\frac{\alpha}{2}\right), \boldsymbol{\Theta}_{\hat{S}_s}^{-1}\left(1 - \frac{\alpha}{2}\right)\right) \tag{26}$$

$$\hat{S}_f \in \left(\boldsymbol{\Theta}_{\hat{S}_f}^{-1}\left(\frac{\alpha}{2}\right), \boldsymbol{\Theta}_{\hat{S}_f}^{-1}\left(1 - \frac{\alpha}{2}\right)\right) \tag{27}$$

where $\boldsymbol{\Theta}_{\hat{S}_s}^{-1}$ and $\boldsymbol{\Theta}_{\hat{S}_f}^{-1}$ are the inverse CDF of the Poisson binomial distribution. According to (23), the upper and lower bounds of the total number of failure points can be derived as:

$$N_f \in \left[\hat{N}_f - \hat{S}_f^u, \ \hat{N}_f + \hat{S}_s^u\right] \tag{28}$$

where $\hat{S}_f^u$ and $\hat{S}_s^u$ are the upper bounds of $\hat{S}_f$ and $\hat{S}_s$, respectively, thus, the maximum error can be estimated as:

$$\epsilon = \left|\frac{\hat{N}_f}{N_f} - 1\right| \leq max\left(\left|\frac{\hat{N}_f}{\hat{N}_f - \hat{S}_f^u} - 1\right|, \left|\frac{\hat{N}_f}{\hat{N}_f + \hat{S}_s^u} - 1\right|\right) = \hat{\epsilon}_{max} \tag{29}$$

where $\hat{S}_f^u = \boldsymbol{\Theta}_{\hat{S}_f}^{-1}\left(1 - \frac{\alpha}{2}\right)$ and $\hat{S}_s^u = \boldsymbol{\Theta}_{\hat{S}_s}^{-1}\left(1 - \frac{\alpha}{2}\right)$. More details for the computation of $\hat{\epsilon}_{max}$ can be found in [35]. Once the maximum error is estimated, the stopping criterion is set as $\hat{\epsilon}_{max} \leq \epsilon_{thr}$. It means that the adaptive training process of the Kriging model stops when $\hat{\epsilon}_{max}$ reaches the prescribed threshold $\epsilon_{thr}$. With the help of ESC, the adaptive Kriging method constructs Kriging models more efficiently while the accuracy of the models is ensured. The adaptive Kriging method with ESC can be a good fit in the VoI analysis. The proposed framework that integrates the adaptive Kriging method with ESC into the VoI analysis is presented in the following section.

## IV. VALUE OF INFORMATION ANALYSIS WITH ADAPTIVE KRIGING

As noted, a primary challenge with VoI analysis lies in the very high computational cost of numerous model evaluations. An adaptive Kriging surrogate modeling with ESC is applied here to construct surrogate models to ease the burden of computational demand. It should be noted that the framework, albeit being presented in the form of a structural reliability method, can be applied to any system or problem where information collection is an option to aid decision making. In order to facilitate the integration of adaptive Kriging into VoI analysis, the framework uses two novel features, which are introduced in detail next.

### A. MODELING SYSTEM RESPONSE

The limit state function for the onset of event *i* can be described as:
$$g_i(\boldsymbol{X}) = Res_{critical,i} - Res_i(\boldsymbol{X}) \tag{30}$$
where $Res_{critical,i}$ is a constant that denotes the critical value of system response for event *i* to occur, e.g., critical deflection of a mid-point of a bridge. In addition, $Res_i(\boldsymbol{X})$ is the response of the system with variables $\boldsymbol{X}$. In adaptive Kriging reliability methods, the limit state function $g_i(\boldsymbol{x})$ is approximated by a Kriging surrogate model. The objective is to improve the accuracy of reliability estimates by training the Kriging model using design points in the vicinity of the limit state. In the VoI analysis, observations **y**, which are sampled based on $\boldsymbol{X}$, are needed for the evaluation of the VoI. The target of data collection efforts is often $Res_i(\boldsymbol{X})$, as it can directly reflect the state of the system. This serves as one of the motivations here to target $Res_i(\boldsymbol{X})$ instead of $g_i(\boldsymbol{X})$ for representation using Kriging surrogate models. The value of $g_i(\boldsymbol{X})$ can be obtained using (30). Note that this scheme can be used for any kind of observation including the system response and random variables describing the properties of the system. In order to ensure the accuracy of the surrogate model in terms of failure probability estimation, the next design point for the limit state function associated with event *i* is selected using $U'$ learning function. This function which is inspired by the original $U'$ learning function in [26] has the following form:
$$U'_i(\boldsymbol{x}_g) = \frac{|Res_{critical,i} - \mu_{Res_i}(\boldsymbol{x}_g)|}{\sigma_{Res_i}(\boldsymbol{x}_g)} \tag{31}$$
where $\mu_{Res_i}(\boldsymbol{x}_g)$ is the estimated system response with respect to $\boldsymbol{x}_g$ by the Kriging model and $\sigma_{Res_i}(\boldsymbol{x}_g)$ is the standard deviation of the estimated system response. Constructing the surrogate model for system response rather than the limit state function allows the implementation of knowledge sharing schemes, which are introduced next, in addition to facilitating identification of the next best design points for improving the accuracy of the surrogate model.

### B. KNOWLEDGE SHARING AMONG SURROGATE MODELS

For different events of interest, especially different damage states, $E_i$, in VoI analysis, a naïve strategy is to model the associated limit state functions separately. However, different limit states concerning different damage states may involve the same system response type. In other words, $Res_i(\boldsymbol{X})$ may be the same for all $E_i$s, and only the critical value of system response varies for different limit states. There can also be cases where groups of events, not all of them, share the same $Res_i(\boldsymbol{X})$. This feature is leveraged here by modeling $Res_i(\boldsymbol{X})$ instead of $g_i(\boldsymbol{X})$, which allows sharing knowledge of the true model among surrogate models for different limit states.

Assume one needs to construct surrogate models for two limit state functions: $g_1(\boldsymbol{X})$ and $g_2(\boldsymbol{X})$, and sharing knowledge between the two surrogate models is desired. Upon the construction of the surrogate model $\widehat{Res}_1(\boldsymbol{X})$, all the design points $\boldsymbol{X}_{g1}$ and the corresponding system responses $\widehat{Res}_1(\boldsymbol{X}_{g1})$ from the construction of the first surrogate are available. Since $Res_1(\boldsymbol{X})$ and $Res_2(\boldsymbol{X})$ both represent the same system response, one can use $\boldsymbol{X}_{g1}$ and $\widehat{Res}_1(\boldsymbol{X}_{g1})$, to generate a sample set and corresponding function values for $\hat{g}_2(X)$ using $\hat{g}_2(X_{g1}) = Res_{critical,1} - \widehat{Res}_1(X_{g1})$. These points and the corresponding responses, albeit provide knowledge about system response in the vicinity of limit state $g_1(X)$, can be leveraged in the construction of $\widehat{Res}_2(X)$, as the same system response is shared. In this sharing scheme, one can take some or all the points from $X_{g1}$ as the initial design points for the construction of $\hat{g}_2(X)$. As shown later in the results section, it is found that as the number of points shared from $X_{g1}$ increases, on average, fewer additional design points are required for the construction of $\hat{g}_2(X)$.

Sharing all design points from the first surrogate model is not the limit for sharing knowledge. The degree of sharing knowledge can be further extended by training the surrogate model concurrently for both limit states. This is possible through targeting the system response for surrogate modeling, as the same response is shared by both limit states. During the process of training the shared surrogate model, the associated $U'$ learning functions must be evaluated for all candidate design points for both limit states. Here it is proposed to compare the two maxima of the $U'$ learning functions for the two limit states and select the candidate design point with the smaller $U'$ learning function as the next training point. Note that once the error for one limit state reaches the specified error threshold estimated using ESC, the surrogate model will be considered as accurate for that limit state and the next training points are selected based on the $U'$ learning function for the remaining limit state. This approach ensures the high efficiency of developing surrogate models and their accuracy. A similar process applies to the check of the coefficient of variation of the probabilities of events, e.g., failure. Once the coefficient of variation of the probability of interest for one limit state has met the requirement, one only needs to check the error and the coefficient of variation of probability for the remaining limit state.

However, sharing model for different limit states is computationally efficient for limit state functions with similar probabilities of events. For limit states that have a large difference in event probabilities, the required number of candidate design points will be substantially different. A relatively small set of candidate design points is required for a limit state function with a relatively large event probability; however, this number of samples may not be sufficient to train a surrogate model for a smaller event probability. On the other hand, a relatively large set of candidate design points required for a limit state function with a relatively small event probability can lead to a very low convergence rate for a limit state with large event probability. When the difference in event probabilities is large, the sharing model scheme may not be applicable, however, knowledge can still be shared among different limit states by sharing the training points.

The two knowledge sharing schemes including sharing model and sharing points can be used together for problems with multiple limit state functions that have a wide range of event probabilities. For such problems, the limit state functions should be grouped first. A shared surrogate model is constructed for each group. All the design points and corresponding limit state function values from group(s) whose shared model has been constructed are passed towards the construction of the shared model for the next group as the initial design points. Note that a higher priority of construction should be given to limit state functions with smaller probabilities of failure, i.e., train models with smaller probability of failure first, as they often require more training points, which can provide more knowledge for the construction of other models.

### C. VOI ANALYSIS WITH ADAPTIVE KRIGING FRAMEWORK
In this section, the proposed framework for calculating VoI using adaptive Kriging is presented. This framework integrates structural reliability methods into VoI analysis based on an adaptive Kriging approach with a new set of attributes that were introduced in the previous subsections. A flowchart illustrating the process is presented in Fig. 1. The main steps of the framework are described below:

- **Step 1:** *Grouping the limit state functions.* Group the limit state functions for different events, e.g. damage states, based on the ranges of associated occurrence probabilities.
- **Step 2**: *Generation of initial candidate design samples.* Generate $N_{MCS}$ candidate design samples using Latin Hypercube Sampling (*LHS*); these samples are denoted as $S$.
- **Step 3**: *Selection of initial training points*. Randomly select from $S$ an initial set of training points denoted as $x_{tr}$ for Kriging construction and evaluate their responses $Res(X_{tr})$.
- **Step 4**: *Select the group with the lowest event probabilities.* Pick the group of limit state functions with the smallest occurrence probabilities. This group will be used to construct the shared surrogate model first. Set error indicators $EI_i = 1$ for $i = 1, …, l$, where l is the number of limit state functions in the group.
- **Step 5**: *Construction of a shared Kriging model for the current group.* Construct the shared Kriging model for the selected group using $X_{tr}$. Denote the Kriging model as $\widehat{Res}(x)$. UQLab [43] with ordinary

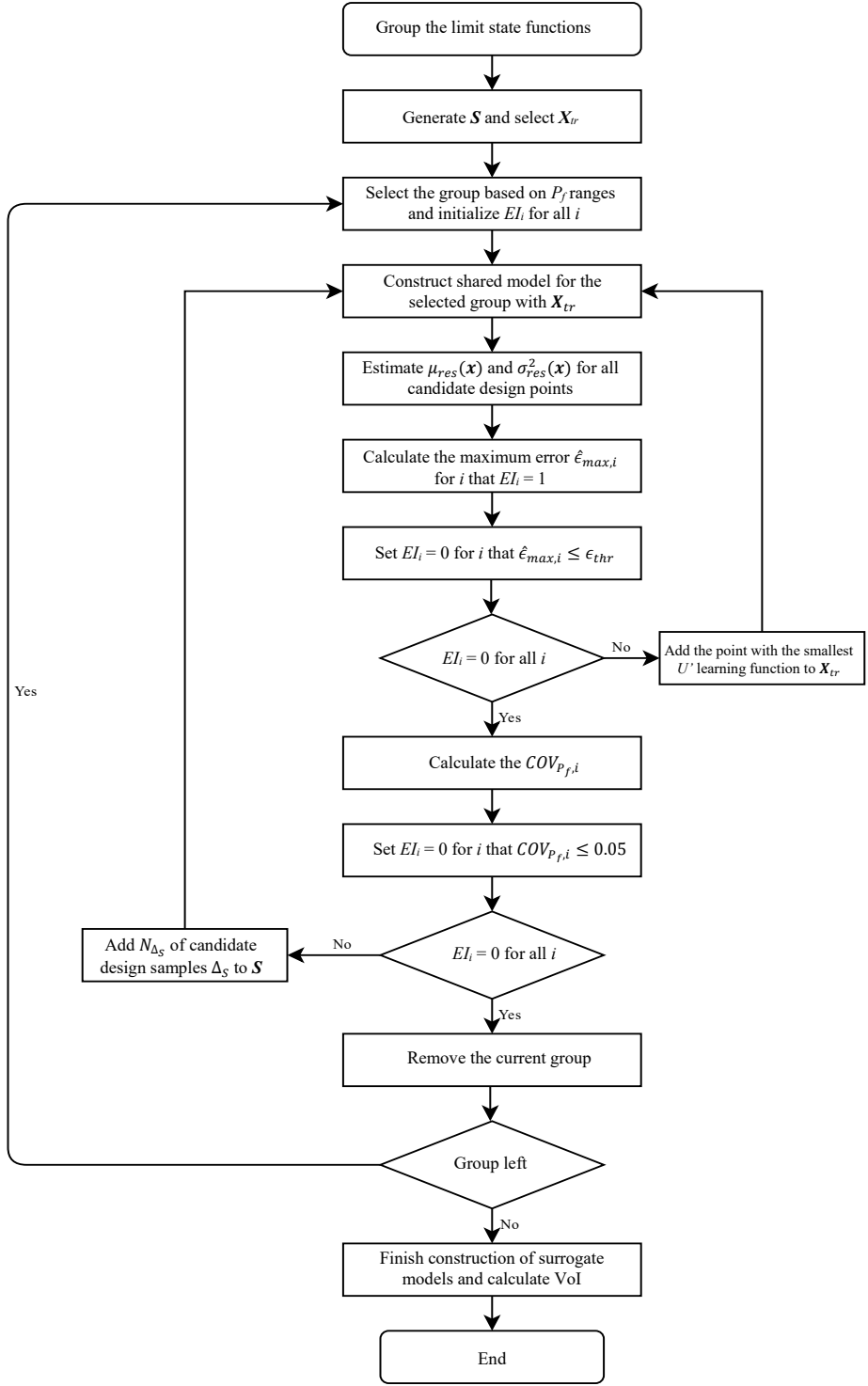

**FIGURE 1.** Proposed framework flowchart

Kriging basis and Gaussian correlation function is used for the construction of the Kriging model.
- **Step 6:** *Kriging estimation.* Obtain the current Kriging responses including the mean $\mu_{res}(x)$ and variance $\sigma_{res}^2(x)$ for all candidate design points. Calculate the $U'$ learning function using (31) with the obtained mean and variance values.
- **Step 7:** *Estimation of the maximum error.* Check the value of $EI_i$. If $EI_i = 1$, calculate the maximum error

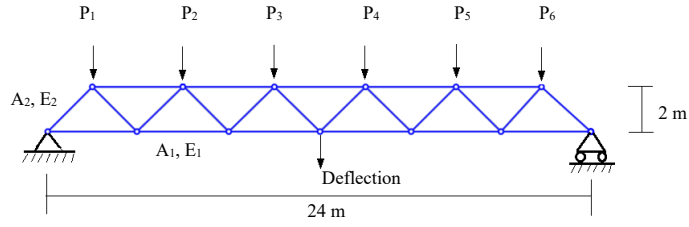

**FIGURE 2. Truss bridge in the example**

$\hat{\epsilon}_{max,i}$ for limit state function *i* using (24)-(29), otherwise, skip the error calculation for limit state function *i* and the error from the last iteration will be used in future steps.
- **Step 8**: *Evaluation of the stopping criterion.* Check the stopping criterion ($\hat{\epsilon}_{max,i} \leq \epsilon_{thr}$). If the stopping criterion is satisfied, then set $EI_i = 0$, otherwise, set $EI_i = 1$. After checking all the stopping criteria for all limit state functions in the group, if $EI_i = 0$ for $i = 1, \ldots, l$, go to Step 10, otherwise, go to Step 9.
- **Step 9**: *Identification of the next training point and updating the training set.* Find the point with the smallest $U'$ value for each limit state function *i* with $EI_i = 1$, evaluate the corresponding $Res(x)$, and then add this point to $X_{tr}$. Afterwards, go to Step 5.
- **Step 10**: *Evaluation of the sufficiency of initial design sample set.* Set error indicators $EI_i = 1$ for $i = 1, \ldots, l$. Then, calculate $\hat{P}_{f,i}$ for each limit state $i$ using Monte Carlo Simulation with the constructed surrogate model. Determine the coefficient of variation of $\hat{P}_{f,i}$ and check the following criterion:

$$COV_{P_f,i} = \sqrt{\frac{1 - \hat{P}_{f,i}}{\hat{P}_{f,i} N_{MCS}}} \leq COV_{thr} \tag{32}$$

where $COV_{\text{thr}}$ is the threshold for the coefficient of variation of $\hat{P}_{f,i}$, and is usually assigned 0.05 [26]. If (32) is satisfied, then set $EI_i = 0$. If not, it means that the number of candidate design samples $N_{MCS}$ is not sufficient, in which case, set $EI_i = 1$ and add an additional number $N_{\Delta_S}$ of candidate design samples $\Delta_S$ to the training set **S**. Afterwards, if $EI_i = 0$ for $i = 1, \ldots, l$, go to Step 11, otherwise, go to Step 5.
- **Step 11**: *Moving to the next group.* The shared surrogate model for the current group is constructed. Use all the training points as the initial sample points for the next group. If there is no more group left, go to Step 12, otherwise go to Step 4.
- **Step 12**: *Calculating the VoI with the constructed surrogate models.* Once all the required surrogate models have been constructed, calculate the VoI using (13) with all the event probabilities and indicator functions evaluated using the surrogate models.

This framework uses Kriging surrogate model with two concurrent knowledge sharing schemes. The proposed adaptive surrogate modeling can substantially improve the efficiency of VoI analysis with significantly fewer model evaluations and achieve the desired accuracy by adopting ESC.

## V. NUMERICAL EXAMPLE
The efficiency and accuracy of the proposed framework is examined in this section for a problem involving a decision for collection of additional information. For this purpose, the benefit of information collection, i.e. VoI, is determined using the proposed method. The calculated VoI is then compared with the cost of information collection, thus, assisting with the decision of whether or not to collect new information. The problem description is given in the following subsection.

### A. PROBLEM STATEMENT
The structure in this example is a bridge adopted from [44], [45]. The bridge is modeled as a truss structure as shown in Fig. 2. Ten independent input random variables are considered, namely the Young's moduli and the cross-section areas of the horizontal and diagonal bars (denoted by $E_1$; $A_1$ and $E_2$; $A_2$, respectively) and the applied loads (denoted by $P_i$, $i = 1, \ldots, 6$). The mean and standard deviation of these variables are reported in Table I. All the random variables are stored in a vector $X = [P_1, P_2, P_3, P_4, P_5, P_6, A_1, A_2, E_1, E_2]$.

The system response of interest is the deflection at mid-span *V*. This response is used to define two events of interest each describing a specific damage state: (1) serviceability failure; and (2) structural failure. For serviceability failure, the deflection limit is considered as 0.09 m. The structural failure corresponds to the situation where the stress in a member exceeds the yield stress. The yield stress considered for the members in the structure is $3.5 \cdot 10^{11}$ Pa. Based on a series of structural analyses, the deflection limit associated with failure is found as 0.11 m. The limit state functions for the serviceability and failure damage states are shown below, respectively:

$$g_{ser}(X) = 0.09 - Res(X) \tag{33}$$

$$g_{str}(X) = 0.11 - Res(X) \tag{34}$$

where $Res(X)$ is the function representing the system response, here, the mid-span deflection.

The following cost model is applied:

1) When the serviceability failure occurs, the bridge will need repair. The sum of the repair cost, the incurred user cost, and the cost of potential injury is estimated as $220,000, which is denoted as $C_{fser}$.

2) When the structural failure occurs, the bridge needs to be replaced. The sum of the cost of replacing the bridge, the incurred user cost, and the cost of potential injury and fatality is estimated as $920,000, which is denoted as $C_{fstr}$.

The following preventive action alternatives are considered for the bridge:

1) $a_0$: do nothing.

2) $a_{per}$: perform a rehabilitation, which is an action that if taken the structure will not fail. The sun of the cost of the rehabilitation and the incurred user costs is estimated as $400,000, which is denoted as $C_{rper}$.

3) $a_{imper}$: perform an imperfect upgrade. In this strategy, members are strengthened with steel plates. The cost of this action is $120,000, which is estimated based on the cost of material and the incurred user cost. The cross-sectional area of the added plates follows a lognormal distribution with a mean of $10^{-4}$ m$^2$ and a standard deviation of $10^{-5}$ m$^2$. Let $X'$ denote the new vector of random variables that is used by the limit state functions corresponding to the case where imperfect upgrade is implemented. This new vector is the same as the original one except that the new variable is added to $A_1$ and $A_2$. In this manner, the dimension of the problem remains the same, and the same surrogate model for the system response can be used for the estimation of the values of limit state functions corresponding to the case where imperfect upgrade is implemented. For the system after the imperfect upgrade, the limit state functions for the two damage states are shown below:

$$g_{ser\prime}(X') = 0.09 - Res(X') \tag{35}$$

$$g_{str\prime}(X') = 0.11 - Res(X') \tag{36}$$

Based on the aforementioned four limit state functions, the relevant events, which determine the consequences and costs, are listed below:

$$\begin{aligned}
E_1 &= \{g_{ser}(X) > 0\} \cap \{g_{ser\prime}(X') > 0\} \\
E_2 &= \{g_{ser}(X) > 0\} \cap \{g_{ser\prime}(X') \leq 0\} \cap \{g_{str\prime}(X') > 0\} \\
E_3 &= \{g_{ser}(X) > 0\} \cap \{g_{str\prime}(X') \leq 0\} \\
E_4 &= \{g_{ser}(X) \leq 0\} \cap \{g_{str}(X) > 0\} \cap \{g_{ser\prime}(X') > 0\} \\
E_5 &= \{g_{ser}(X) \leq 0\} \cap \{g_{str}(X) > 0\} \cap \{g_{ser\prime}(X') \leq 0\} \cap \{g_{str\prime}(x') > 0\} \quad (37) \\
E_6 &= \{g_{ser}(X) \leq 0\} \cap \{g_{str}(X) > 0\} \cap \{g_{str\prime}(X') \leq 0\} \\
E_7 &= \{g_{str}(X) \leq 0\} \cap \{g_{ser\prime}(X') > 0\} \\
E_8 &= \{g_{str}(X) \leq 0\} \cap \{g_{ser\prime}(X') \leq 0\} \cap \{g_{str\prime}(X') > 0\} \\
E_9 &= \{g_{str}(X) \leq 0\} \cap \{g_{str\prime}(X') \leq 0\}
\end{aligned}$$

Note that $g_{ser}(X)$ and $g_{str}(X)$ are always larger than $g_{ser\prime(X)}$ and $g_{str\prime}(X)$, respectively. These nine events can be further grouped together into 6 events as follows:

TABLE I
UNITS FOR MAGNETIC RANDOM VARIABLE PARAMETERS PROPERTIES

| Variable | Distribution | Mean | St. dev. |
|---|---|---|---|
| $E_1$, $E_2$ (Pa) | Lognormal | $2.1 \cdot 10^{11}$ | $2.1 \cdot 10^{10}$ |
| $A_1$ (m$^2$) | Lognormal | $4.0 \cdot 10^{-3}$ | $2.0 \cdot 10^{-3}$ |
| $A_2$ (m$^2$) | Lognormal | $3.0 \cdot 10^{-3}$ | $1.5 \cdot 10^{-3}$ |
| $P_1$-$P_6$ (N) | Gumbel | $4.0 \cdot 10^4$ | $4. \cdot 10^{-4}$ |

$$E_1' = \{g_{ser}(X) > 0\}$$
$$E_2' = \{g_{ser}(X) \leq 0\} \cap \{g_{str}(X) > 0\} \cap \{g_{ser\prime}(X') > 0\}$$
$$E_3' = \{g_{ser}(X) \leq 0\} \cap \{g_{str}(X) > 0\} \cap \{g_{ser\prime}(X') \leq 0\} \cap \{g_{str\prime}(X') > 0\} \quad (38)$$
$$E_4' = \{g_{str}(X) \leq 0\} \cap \{g_{ser\prime}(X') > 0\}$$
$$E_5' = \{g_{str}(X) \leq 0\} \cap \{g_{ser\prime}(X') \leq 0\} \cap \{g_{str\prime}(X') > 0\}$$
$$E_5' = \{g_{str}(X) \leq 0\} \cap \{g_{str\prime}(X') \leq 0\}$$

The costs associated with these six mutually exclusive events and the actions taken in response to the events are listed in Table II.

The information collection action in this example pertains measuring the mid-span displacement of the bridge in a load test. This measurement is the system response of the structure, which is also the target output for the Kriging models. Let $P$ denote the random variable vector containing the external load random variables
(i.e., $P_1, P_2, …, P_6$) and $M$ denote the random variable vector containing the geometric and material property variables (i.e., $E_1, E_2, A_1$ and $A_2$). When the load testing is performed, all applied loads are fixed to $4.0 \cdot 10^4$ N; therefore, only random variables in $M$ will remain uncertain during the load test. Following Bayes' rule, the conditional distribution of $M$ given an observation $Z$ (i.e., the observed mid-span deflection in the load testing) can be formulated as:

$$f_{m|z}(m) = \frac{L_m(m)f_m(m)}{\int_m L_m(m)f_m(m)dm} \quad (39)$$

The likelihood function $L_m(m)$ of the measurement outcome $y_m$ describes the relation between $Z$ and $M$. Following (8), $L_m(m)$ can be expressed as:

$$L_m(m) = f_{Y_m}(y_m|m) = f_\varepsilon[y_m - Res'(m)] \quad (40)$$

where $Res'(m)$ is the mid-span deflection when $M = m$ and $P = 4.0 \cdot 10^4$ N, and $f_\varepsilon$ is the normal probability density function with a mean of 0 and standard deviation of
ε (the measurement error). Using MCS, the posterior probability of an event $E_i$ can be determined as:

$$\begin{aligned}
Pr(E_i|Z) &= \int_x I(m, p \in \Omega_{E_i}) f_{m|z}(m) f_p(p) dmdp \\
&= \int_x I(m, p \in \Omega_{E_i}) \frac{L_m(m)f_m(m)}{\int_m L_m(m)f_m(m)dm} f_p(p) dmdp \\
&= \frac{\int_x I(m, p \in \Omega_{E_i}) L_m(m)f_m(m)f_p(p)dmdp}{\int_m L_m(m)f_m(m)dm} \\
&\approx \frac{\sum_{k=1}^{n_{MCS}} I(m_k, p_k \in \Omega_{E_i}) L_m(m_k)}{\sum_{k=1}^{n_{MCS}} L_m(m_k)}
\end{aligned} \quad (41)$$

where $f_p(p)$ and $f_m(m)$ are the prior joint probability density function of $P$ and $M$, respectively. Moreover, $m_k, p_k, k = 1, …, n_{MCS}$, are samples from $f_m(m)$ and $f_p(p)$, respectively. Thus, similar to (13) the $VoI$ can

TABLE II
COSTS ASSOCIATED WITH EVENTS AND ACTIONS

| Event | $a_0$ | $a_{per}$ | $a_{inp}$ |
|---|---|---|---|
| $E_1'$ | 0 | $C_{rper}$ | $C_{rimp}$ |
| $E_2'$ | $C_{fser}$ | $C_{rper}$ | $C_{rimp}$ |
| $E_3'$ | $C_{fser}$ | $C_{rper}$ | $C_{rimp} + C_{fser}$ |
| $E_4'$ | $C_{fser}$ | $C_{rper}$ | $C_{rimp}$ |
| $E_5'$ | $C_{fser}$ | $C_{rper}$ | $C_{rimp} + C_{fstr}$ |
| $E_6'$ | $C_{fser}$ | $C_{rper}$ | $C_{rimp} + C_{fstr}$ |

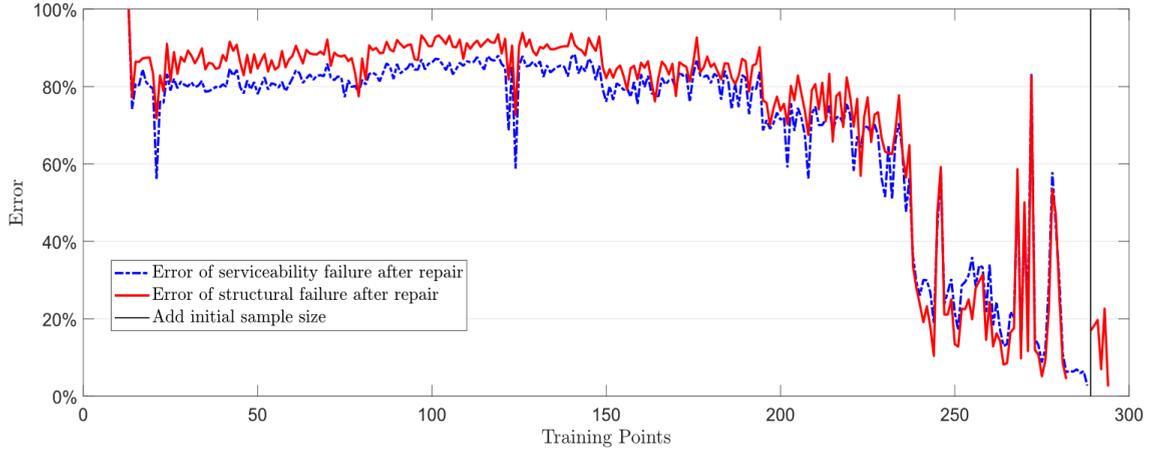

**FIGURE 3.** Error Trends of Kriging surrogate models

be determined as:

$$VoI \approx C_{prior} - \frac{1}{n_{SY}} \times \sum_{j=1}^{n_{SY}} \min_{a} \sum_{i=1}^{m} C_{E_i}(a) \frac{\sum_{k=1}^{n_{MCS}} I(\boldsymbol{m}_k, \boldsymbol{p}_k \in \Omega_{E_i}) L_m(\boldsymbol{m}_k)}{\sum_{k=1}^{n_{MCS}} L_m(\boldsymbol{m}_k)} \quad (42)$$

### B. SURROGATE MODEL CONSTRUCTION

In this example, surrogate models are constructed for the system response $\boldsymbol{Res}(\boldsymbol{X})$. The four limit state functions are first divided into two groups: the two limit state functions for the system before upgrading, $\boldsymbol{g}_{ser}(\boldsymbol{X})$ and $\boldsymbol{g}_{str}(\boldsymbol{X})$, are in one group, and the two remaining limit states, $\boldsymbol{g}_{ser\prime}(\boldsymbol{X}')$ and $\boldsymbol{g}_{str\prime}(\boldsymbol{X}')$, are in another group. This grouping scheme, which is based on the failure probabilities associated with the limit states, allows both model sharing and sharing points schemes to work here. The shared model of the latter group is constructed first as the probabilities of failure for the system after upgrade are smaller than those before the upgrade. These training points will provide knowledge about system response for the next shared model. The initial sample size for the first group is $10^5$, while the one for the second group is $10^4$. Fig. 3 shows the error trend of one realization for the first group. As shown in the figure, the error for both limit states are evaluated in each iteration. When the error of the limit state for structural failure after the upgrade reaches the threshold of 5%, the model is considered accurate for that limit state. When both errors are smaller than the threshold, COVs of the probabilities of failure for both limit states are checked. As the COV of the probability of failure for structural failure after the upgrade does not satisfy the corresponding criterion, the initial sample size is increased by $N_{\Delta_S}$ of $10^5$. Additional training points are required to meet the error threshold for the limit state corresponding to structural failure after the upgrade. Afterwards, all requirements are met and the construction of the shared model for the first group is complete. The construction of the shared model for the second group begins with all the training points for the first group as the initial training points.

The constructed models are used to evaluate the probabilities of the six events using MCS. The last surrogate model with the most training points is used to estimate the
system response for the calculation of VoI.

### C. NUMERICAL RESULTS

After the construction of the surrogate models, the VoI can be calculated using (42) with all the probabilities of failure and indicator functions evaluated using the surrogate models. The results evaluated by repeating the computations 20 times are shown in Table III. With only 321 model evaluations on average, the proposed framework is able to yield an accurate estimate of the VoI with 2.5% error compared to MCS results. Moreover, the standard deviation of the VoI estimated using the proposed method is small indicating the robustness of this method. The MCS approach, on the other hand, requires $10^6$ model evaluations for probability of failure calculation and $10^4$ model evaluations for the calculation of VoI. Note that the $10^4$ model evaluations can be selected from the $10^6$ model evaluations to reduce the computational demand. However, this number is still quite large.

The IS-based scheme proposed by Straub [17] highly depends on MLFPs. In fact, the construction of IS density function and the calculation of the probability of failure depend on MLFPs. In this study, the IS-based scheme was applied to the bridge problem, however, accurate MLFPs for analyzing the reliability of the system are not available. Therefore, the approach in [17] cannot solve the VoI analysis problem in this example.

In addition, the application of Kriging-based reliability methods such as AK-MCS [26] was explored here. However, the results were highly inaccurate. This limitation is due to the inappropriate stoppage of the training process of surrogate models by learning functions such as $U'$ and EFF. In the bridge example, $U'$ required stoppage of the training process following the initial twelve training points, the estimates of failure probabilities by the resulting Kriging model at this stage was zero. With the estimate using $U'$, the stopping criterion of the next stage in AK-MCS which is based on the COV of failure probability is not satisfied, leading to an increase in the candidate design sample size. With this new sample set, $U'$ again indicate stoppage of the training process following the initial training samples and yielding probability estimate of zero. This loop continues with AK-MCS without any sign of improvement. On the other hand, EFF didn't shown any signs of convergence for a single limit state even with 500 training points. However, ESC offers a significantly more robust stopping criterion and approach to VoI analysis based on adaptive surrogate models.

It should be noted that without the proposed knowledge sharing schemes, even using ESC as the stopping criterion, the number of model evaluations can easily exceed 1,000. The knowledge sharing schemes greatly improve the efficiency of the VoI analysis.

As shown in Table III, the estimated VoI is $1629 by MCS and $1588 by the proposed method. Assuming that the load testing costs $2000, the VoI for this measurement is less than the cost of the information collection. Therefore, the optimal decision is not to perform the load testing. It should be noted that as infrastructure systems age, uncertainties in their capacity to withstand stressors increase. Load testing in such circumstances can yield more benefits compared to cases where the extent of uncertainty is limited.

### D. KNOWLEDGE SHARING SCHEME DISCUSSION

The performance of different knowledge sharing schemes is investigated here. Surrogate models are constructed for $g_{ser}(X)$ and $g_{str}(X)$ using different knowledge sharing schemes. Results evaluated by repeating the process 50 times are summarized in Table IV. It is observed that as the number of shared points increases, the total number of training points decreases. When a shared model is used, the number of training points is minimum. This observation indicates that the shared model scheme is able to share the knowledge to the greatest degree and is therefore the most efficient strategy. However, when a large difference in different limit states is present, shared model scheme may not work efficiently and shared points scheme can be utilized. In this bridge example, the failure probabilities regarding $g_{ser}(X)$ and $g_{str}(X)$ are $4.6 \cdot 10^{-2}$ and $3.8 \cdot 10^{-2}$, respectively, and the failure probabilities regarding $g_{ser'}(X')$ and $g_{str'}(X')$ are $5.8 \cdot 10^{-3}$ and $4.4 \cdot 10^{-3}$, respectively. The former two are significantly different from the latter two. Thus, the former two limit state functions can share a model and the latter two limit state functions share another model. In addition, training points are shared between the two shared models. As seen here, both knowledge sharing schemes are leveraged in this example. It should be noted that the knowledge sharing schemes are not applicable for problems with totally different system response functions in the limit states. In such cases, the proposed framework is still applicable by constructing different surrogate models independently.

### VI. CONCLUSIONS

VoI analysis offers the capability to quantify the potential benefits afforded by avoiding errors in decision making through collection of new information. The high computational cost of evaluating models is one of the primary challenges facing VoI analysis. To address this challenge, this paper developed a framework that integrates adaptive Kriging surrogate modeling into VoI analysis. The framework uses two novel knowledge sharing schemes that substantially reduce the number of model evaluations required for the analyses. The proposed method was applied to perform VoI analysis for a bridge structure. This problem could not be solved using a state-of-the-art approach based on Importance Sampling nor using AK-MCS method, which is a capable and popular adaptive Kriging-based reliability analysis approach. The proposed method, on the other hand, was able to accurately and efficiently estimate the VoI. Results from repeating this analysis using the proposed approach indicated the robustness of the framework.

TABLE III
NUMERICAL EXAMPLE RESULTS

| Method | VoI mean | VoI st. dev. | Average number of model evaluations | Range of number of model evaluations |
|---|---|---|---|---|
| MCS | 1629 | 309 | $10^4 + 10^6$ | - |
| Proposed framework | 1588 | 221 | 321 | 268~381 |

TABLE IV
KNOWLEDGE SHARING SCHEME COMPARISON

| Scheme | Number of training points required |
|---|---|
| Separate models | 337 |
| Sharing 50 training points | 276 |
| Sharing all training points | 244 |
| Sharing model | 235 |

The proposed approach can be improved in two respects, both related to ESC. First, the rate of convergence to the desired accuracy can be slow for a problem with a significantly small probability of failure. As the size of training points increases, the computational cost for the construction of surrogate models increases. Improving the accuracy of error estimation in ESC for such problems can further improve the computational efficiency. Secondly, the error control scheme in ESC is focused on the error of the prior probability. However, posterior error estimation can offer a stopping criterion that can further enhance the accuracy of the models and therefore VoI estimates.

**ACKNOWLEDGEMENT**
This research was supported in part by the U.S. National Science Foundation (NSF) through awards CMMI-1635569 and 1762918, and the Lichtenstein endowment. Any opinions, findings, and conclusions or recommendations expressed in this paper are those of the authors and do not necessarily reflect the views of the National Science Foundation.